\documentclass[journal=jacsat,manuscript=article]{achemso}

\usepackage{chemformula} 
\usepackage[T1]{fontenc} 

\usepackage[T1]{fontenc}
\usepackage[latin9]{inputenc}
\usepackage{color}
\usepackage{amsmath}
\usepackage{amssymb}
\usepackage{graphicx}
\usepackage{esint}

\usepackage{babel}

\makeatletter
\@ifundefined{textcolor}{}
{%
 \definecolor{BLACK}{gray}{0}
 \definecolor{WHITE}{gray}{1}
 \definecolor{RED}{rgb}{1,0,0}
 \definecolor{GREEN}{rgb}{0,1,0}
 \definecolor{BLUE}{rgb}{0,0,1}
 \definecolor{CYAN}{cmyk}{1,0,0,0}
 \definecolor{MAGENTA}{cmyk}{0,1,0,0}
 \definecolor{YELLOW}{cmyk}{0,0,1,0}
 }

\makeatother

\newcommand{\p}{\partial}

\newcommand{\calP}{\mathcal{P}}

\newcommand{\bnabla}{\mbox{\boldmath $\nabla$}}

\newcommand{\be}{\begin{equation}}
\newcommand{\ee}{\end{equation}}
\newcommand{\bea}{\begin{eqnarray}}
\newcommand{\eea}{\end{eqnarray}}
\newcommand{\beqa}{\begin{eqnarray*}}
\newcommand{\eeqa}{\end{eqnarray*}}
\newcommand{\nn}{\nonumber}

\newcommand{\ba}{\begin{array}{c}}
\newcommand{\baa}{\begin{array}{cc}}
\newcommand{\baaa}{\begin{array}{ccc}}
\newcommand{\baaaa}{\begin{array}{cccc}}
\newcommand{\ea}{\end{array}}

\newcommand{\bma}{\left[\begin{array}{c}}
\newcommand{\bmaa}{\left[\begin{array}{cc}}
\newcommand{\bmaaa}{\left[\begin{array}{ccc}}
\newcommand{\bmaaaa}{\left[\begin{array}{cccc}}
\newcommand{\ema}{\end{array}\right]}

\author{Jayden Craft$^{(1)}$}

\author{Muhammad Waqas Shabbir$^{(1)}$}

\author{Dirk R. Englund$^{(2)}$}

\author{R. M. Osgood III$^{(3)}$}

\author{Michael N. Leuenberger$^{(1,4)}$}

\email{michael.leuenberger@ucf.edu}

\affiliation{$^{(1)}$ NanoScience Technology Center and Department of Physics, University of Central Florida, Orlando, FL 32826, USA. \\
$^{(2)}$ Department of Electrical Engineering and Computer Science, Massachusetts Institute of Technology, Cambridge, MA 02139, USA.\\
$^{(3)}$ U.S. Army Combat Capabilities Development Command Soldier Center, Natick, MA 01760, USA.\\
$^{(4)}$ College of Optics and Photonics, University of Central Florida, Orlando, FL 32826, USA.}

\title{Spectrally Selective Thermal Emission from Graphene Decorated with Metallic Nanoparticles}

\abbreviations{IR,NMR,UV}
\keywords{American Chemical Society, \LaTeX}

\begin{document}

\begin{tocentry}

Some journals require a graphical entry for the Table of Contents.
This should be laid out ``print ready'' so that the sizing of the
text is correct.

Inside the \texttt{tocentry} environment, the font used is Helvetica
8\,pt, as required by \emph{Journal of the American Chemical
Society}.

The surrounding frame is 9\,cm by 3.5\,cm, which is the maximum
permitted for  \emph{Journal of the American Chemical Society}
graphical table of content entries. The box will not resize if the
content is too big: instead it will overflow the edge of the box.

This box and the associated title will always be printed on a
separate page at the end of the document.

\end{tocentry}

\begin{abstract}
We showed in past work that nanopatterned monolayer graphene (NPG) enables spectrally selective thermal emission in the mid-infrared (mid-IR) from 3 to 12 $\mu$m.
In that case the spectral selection is realized by means of the localized surface plasmon (LSP) resonances inside graphene.
Here we show that graphene decorated with metallic nanoparticles, such as Ag nanocubes or nanospheres, also realize spectrally selective thermal emission, but in this case by means of acoustic graphene plasmons (AGPs) localized between graphene and the Ag nanoparticle inside a dielectric material.
Our finite-difference time domain (FDTD) calculations show that the spectrally selective thermal radiation emission can be tuned by means of a gate voltage into two different wavelength regimes, namely the atmospherically opaque regime between $\lambda=5$ $\mu$m and $\lambda=8$ $\mu$m or the atmospherically transparent regime between $\lambda=8$ $\mu$m and $\lambda=12$ $\mu$m. This allows for electric switching between radiative heat trapping mode for the fomer regime and radiative cooling mode for the latter regime.
Our theoretical results can be used to develop graphene-based thermal management systems for smart fabrics.

\textcolor{black}{KEYWORDS: Graphene, metallic nanoparticles, acoustic graphene plasmons, thermal emission. }
\end{abstract}


\section{Introduction}
Spectrally selective engineering of thermal emission is used widely, from clothing to thermal management of computer chips and batteries or photovoltaics that need to be kept at optimal operating temperatures. In cold temperature environments, low emittance in the entire infrared (IR) wavelength regime reduces heat loss. In warm temperature environments, high emittance in the atmospherically transparent IR regime between $\lambda=8$ $\mu$m and 13 $\mu$m enables radiative cooling to the 3 K outer space temperature. 
For example, human skin has a very large emittance of $\epsilon=0.98$ in the IR wavelength regime between 7 and 14 $\mu$m, with a peak at $\lambda=9.5$ $\mu$m. While this large emittance is desirable in the summer, it is detrimental in the winter.

Advanced textiles for personal thermal management have been recently reviewed.\cite{Peng2020} Textiles focus on evaporation, convection, conduction, and thermal radiation for heat management. Advanced textiles for radiative cooling and warming need to be developed by controlling passively and/or actively the emittance $\epsilon$, transmittance $T$, reflectance $R$, and absorbance $A$, which satisfy Kirchhoff's law of thermal radiation $T+R+A=1$, with $\epsilon=A$ in thermal equilibrium. Maximum radiative cooling is achieved for $\epsilon=1$ or $T=1$, whereas maximum radiative warming is realized for $R=1$. Mid-IR transparent radiative cooling textiles could be made of porous polyethelene (NanoPE) fibers. Mid-IR emissive radiative cooling textiles could consist of a highly emissive outer layer, e.g. made of carbon fibers, and a low emissive inner layer, e.g. made of copper. Inverting this double-layer structure could be used for radiative warming.\cite{Hsu2017}
Daytime radiative cooling has been proposed using dielectric materials and photonic crystals\cite{Rephaeli2013} or a polymer-coated fused silica mirror.\cite{Kou2017}
Highly efficient radiative cooling has been demonstrated using an array of symmetrically shaped conical metamaterial made of Al and Ge layers.\cite{Hossain2015} 
Radiative cooling to sub-freezing temperatures has been demonstrated using layers of 
Si$_3$N$_4$, amorphous silicon, and Al as selective thermal emitter.\cite{Chen2016}
Self-adaptive radiative cooling based on phase change materials has recently been proposed.\cite{Ono2018}
Clothes with Ag nanowires or Ag nanoparticles reflect human IR radiation and can therefore be used for radiative warming.\cite{Peng2020} 
Personal thermal management systems have recently been demonstrated with Kevlar fiber and reduced graphene oxide (rGO) composite materials\cite{Hazarika2018} and with Ag nanowires and rGO composite materials.\cite{Hazarika2019}

Control over broadband IR emission is not only useful for radiative heating and cooling for thermal management, but also for IR camouflage. A recent system based on multilayer graphene has been shown experimentally to exhibit tunable IR emittance through gate-controlled reversible intercalation of ionic liquids.\cite{Salihoglu2018} Spectrally selective thermal emission opens up the opportunity for IR camouflage in the atmospherically opaque window between 5 and 8 $\mu$m wavelengths, which has been demonstrated experimentally in ZnS/Ge multilayers.\cite{Zhu2021}  

In recent years, several methods have been implemented for achieving a spectrally selective emittance, in particular narrowband emittance, which increases the coherence of the emitted  photons. One possibility is to use a material that exhibits optical resonances due to the band structure or due to confinement of the charge carriers.\cite{Baranov2019}
Another method is to use structural optical resonances to enhance and/or suppress the emittance. Recently, photonic crystal structures have been used to implement passive pass band filters that reflect the thermal emission at wavelengths that match the photonic bandgap.\cite{Cornelius1999,Lin2000} 
Alternatively, a truncated photonic crystal can be used to enhance the emittance at resonant frequencies.\cite{Celanovic2005,Yang2017}
Our recent theoretical study reveals that nanopatterned graphene (NPG) can be used for gate-tunable spectrally selective thermal emission in the wavelength regime from 3 to 14 $\mu$m.\cite{Shabbir2020}
Nanopatterning graphene provides a method to increase the absorbance and emittance of pristine graphene from around 2\% to nearly 100\%.\cite{Safaei2017,SafaeiACS,Shabbir2020}
This large absorbance can be used to implement an infrared photodetector based on the photothermoelectric effect.\cite{Safaei2019}
Our recent theoretical proposal for a IR photodetector based on multilayer graphene intercalated with FeCl$_3$ achieves large absorbance and emittance with gate-tunable spectral selectivity down to a wavelength of $\lambda=1.3$ $\mu$m.\cite{Shabbir2022}

Recently, acoustic graphene plasmons (AGPs) at the interface between graphene and Ag nanocubes have been observed, achieving extreme confinement of the electromagnetic field in the IR regime.\cite{Epstein2020} The AGPs are acoustic in the sense that their dispersion relation $\omega_{\rm AGP}(q)\propto q$ is linear in $q$, in contrast to standard graphene plasmons with $\omega_{\rm GP}(q)\propto \sqrt{q}$. 

\begin{figure*}[htb]
\begin{centering}
\includegraphics[width=16.0cm]{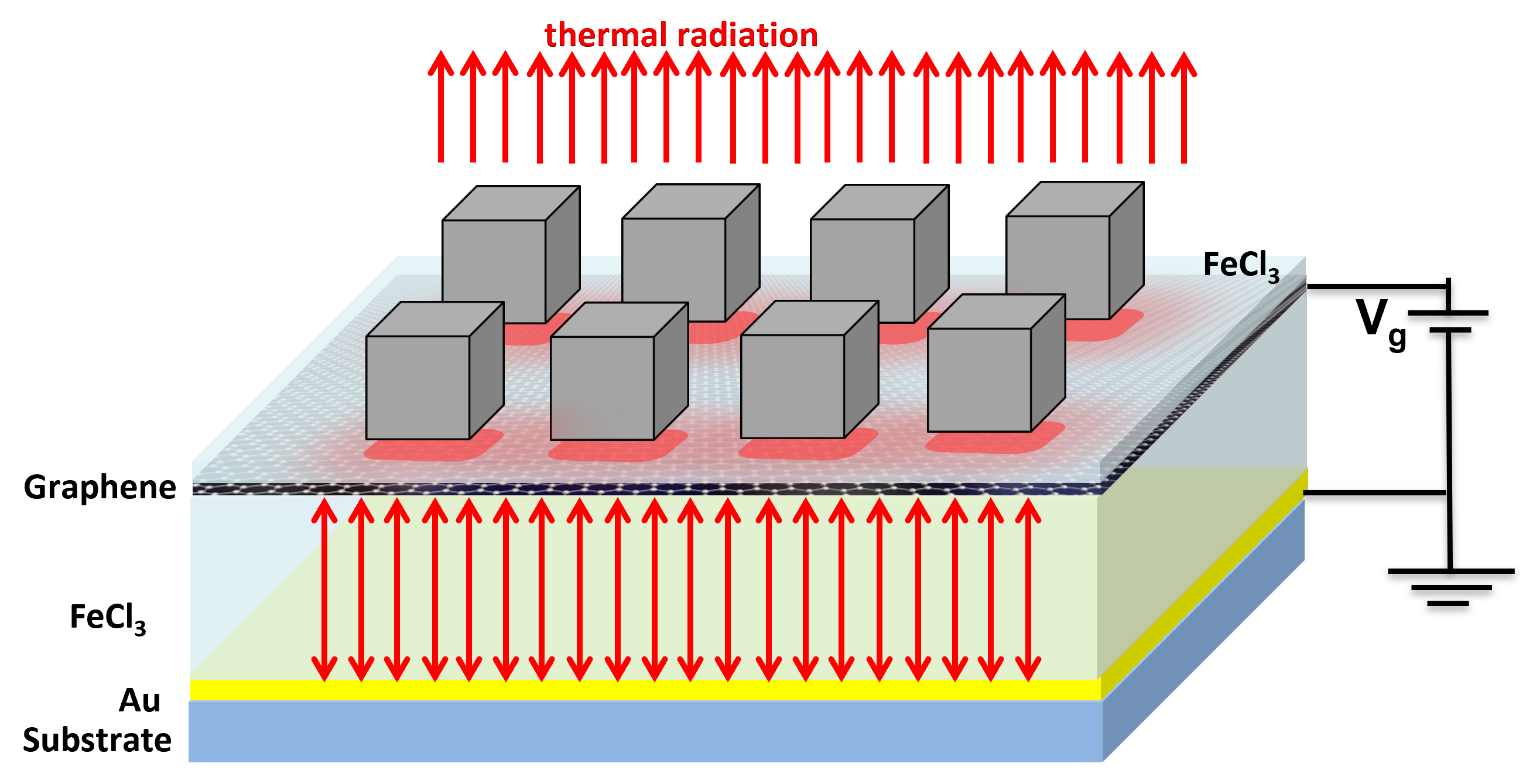}
\end{centering}
\caption{Schematic showing our proposed Ag nanocube/hBN/graphene heterostructure placed on top of a cavity, which can be tuned by means of a gate voltage applied to the Au back mirror. The spacer of the cavity consists of FeCl$_3$.  
\label{fig:AgNanocube-graphene} }
\end{figure*}

Here, we show that gate-tunable AGP resonances can be realized for spectrally selective thermal emission in the atmospherically opaque window between $\lambda=5$ $\mu$m and $\lambda=8$ $\mu$m and also in the atmospherically transparent window between $\lambda=8$ $\mu$m and $\lambda=12$ $\mu$m by means of Ag nanoparticles on top of a dielectric/graphene heterostructure. 
We consider two types of Ag nanoparticles:
Ag nanocubes and Ag nanospheres.
The heterostructure consisting of Ag nanocubes on top of hexagonal boron nitride (hBN)/graphene is shown in Fig.~\ref{fig:AgNanocube-graphene}.
The method to tune the spectrally selective thermal emission in the heterostructure Ag nanoparticle/dielectric/graphene by means of a gate voltage $V_g$ that varies
the Fermi energy $E_F$ inside graphene, thereby varying the charge density and therefore resonance wavelength of the AGPs 
in the wavelength regime between 5 $\mu$m and 12 $\mu$m.

\section{AGP for the nanocube-dielectric-graphene system}
\subsection{Analytical derivations}
Among a variety of possible metallic nanoparticles, it is possible to derive analytical results for nanocubes.
Let us first determine the approximate solution of the AGPs inside the dielectric region sandwiched between a nanocube and graphene.
Let us solve the electromagnetic problem of a classical field inside this cavity. We can find the solutions of the vector potential $A_z$ using the Helmholtz equation
\be
\bnabla^2 A_z = k^2 A_z.
\ee
The TM$^z$ solutions are
\begin{align}
    A_z &= \left[A_1\cos\left(k_x x\right) + B_1\sin\left(k_x x\right)\right]\left[A_2\cos\left(k_y y\right) + B_2\sin\left(k_y y\right)\right] \nn\\
    & \times\left[A_3e^{\zeta_z z} + B_3e^{-\zeta_z z}\right] \nn\\
    &= A_{mn}\cos\left(k_x x\right)\cos\left(k_y y\right)\cosh\left(\zeta_z z\right)
    \label{eq:A_ansatz_graphene}
\end{align}
This ansatz needs to satisfy the boundary conditions of perfect magnetic conductors at the side walls of the cavity, i.e.
\be
    H_x(y=-W/2) = H_x(y=W/2)=0,
    H_y(x=-L/2) = H_y(x=L/2)=0,
    \label{eq:wallboundary}
\ee
where $-h\le z \le h$, $-L/2\le x \le L/2$, $-W/2\le y \le W/2$.
The boundary conditions due to the graphene sheet and its image are
\begin{align}
    E_x(z=-h,z\le -h) &= E_x(z=-h,-h\le z\le h), \nn\\
    H_y(z=-h,z\le -h) &= H_y(z=-h,-h\le z\le h) + \mu_0\sigma(\omega) E_x(z=-h,z\le -h), \label{eq:graphene_boundary}\\
    E_x(z=h,-h\le z\le h) &= E_x(z=h,z\ge h), \nn\\
    H_y(z=h,-h\le z\le h) &= H_y(z=h,z\ge h)  + \mu_0\sigma(\omega) E_x(z=h,z\ge h). \label{eq:graphene_image_boundary}
\end{align}
Note that the nanocube surface is at $z=0$ and the graphene sheet is located at $z=-h$.

Due to the boundary conditions imposed by the perfect magnetic conductors at the side walls [see Eq.~(\ref{eq:wallboundary})] the wavenumbers $k_x$ and $k_y$ become discrete, i.e.
\begin{align}
    k_x &= \left(\frac{m\pi}{L}\right), \, m=0,1,2,\ldots, \nn\\
    k_y &= \left(\frac{n\pi}{W}\right), \, n=0,1,2,\ldots,  .
\end{align}
If $m=n$, then $m=n\ne 0$.
The constraint equation for the wavenumbers is
\be
k_x^2+k_y^2-\zeta_z^2 = k_r^2 = \omega_r^2\mu\epsilon = \epsilon_r\omega_r^2/c^2 \ge 0
\ee
Note that the $\zeta_z^2$ term is negative because the $k_z$ vector is purely imaginary, i.e. $k_z=i\zeta_z$.
Thus, the resonant frequencies of the cavity are
\begin{align}
    f_r &= \frac{1}{2\pi\sqrt{\mu\epsilon}}\sqrt{\left(\frac{m\pi}{L}\right)^2+\left(\frac{n\pi}{W}\right)^2-\zeta_z^2} \nn\\
    &= \frac{c}{2\pi\sqrt{\epsilon_r}}\sqrt{\left(\frac{m\pi}{L}\right)^2+\left(\frac{n\pi}{W}\right)^2-\zeta_z^2},
\end{align}
where $\epsilon_r$ is the relative permittivity of the material inside the cavity and $c$ is the speed of light in vacuum. The resonant wavelengths are $\lambda_r=c/f_r$.
Therefore, the electric and magnetic fields within the cavity are
\begin{align}
    E_z &= -i\frac{1}{\omega\mu\epsilon} \left(\frac{\p^2}{\p z^2}+k^2\right)A_z 
    =-i\frac{\left(k^2+\zeta_z^2\right)}{\omega\mu\epsilon}A_{mn}\sin\left(k_x x\right)\sin\left(k_y y\right)\cosh\left(\zeta_z z\right) \nn\\
    &= -i\frac{\left(k_x^2+k_y^2\right)}{\omega\mu\epsilon}A_{mn}\sin\left(k_x x\right)\sin\left(k_y y\right)\cosh\left(\zeta_z z\right) \nn\\
    E_x &= -i\frac{1}{\omega\mu\epsilon} \frac{\p^2 A_z}{\p z \p x}
    =i\frac{\zeta_z k_x}{\omega\mu\epsilon}A_{mn}\cos\left(k_x x\right)\sin\left(k_y y\right)\sinh\left(\zeta_z z\right), \nn\\
    E_y &= -i\frac{1}{\omega\mu\epsilon} \frac{\p^2 A_z}{\p z \p y}
    =i\frac{\zeta_z k_y}{\omega\mu\epsilon}A_{mn}\sin\left(k_x x\right)\cos\left(k_y y\right)\sinh\left(\zeta_z z\right), \nn\\
    H_z &= 0 \nn\\
    H_x &= \frac{1}{\mu} \frac{\p A_z}{\p y}
    =\frac{k_y}{\mu}A_{mn}\sin\left(k_x x\right)\cos\left(k_y y\right)\cosh\left(\zeta_z z\right), \nn\\
    H_y &= -\frac{1}{\mu} \frac{\p A_z}{\p x}
    =-\frac{k_x}{\mu}A_{mn}\cos\left(k_x x\right)\sin\left(k_y y\right)\cosh\left(\zeta_z z\right).
\end{align}

\begin{figure*}[htb]
\begin{centering}
\includegraphics[width=16cm]{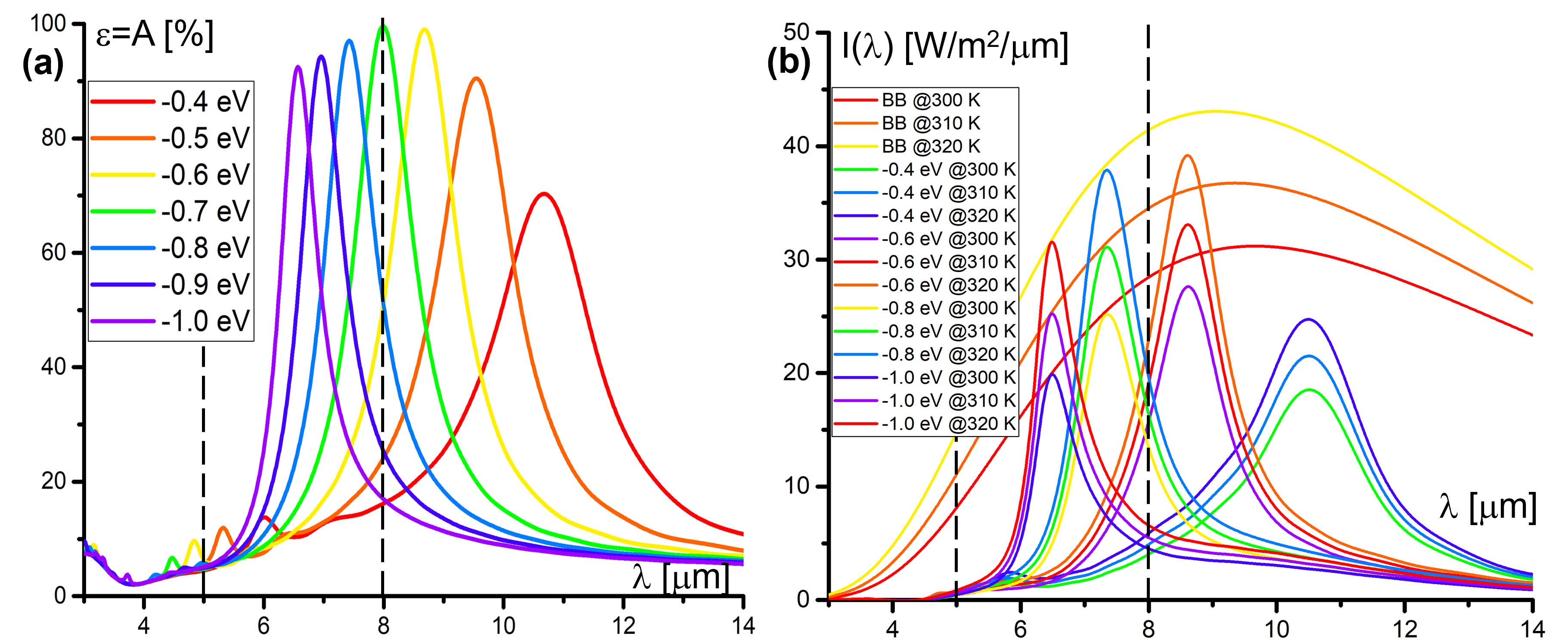}
\end{centering}
\caption{(a) Emittance $\epsilon(\lambda)$ (=absorbance $A(\lambda)$) and (b) spectral radiance $I(\lambda)$ as functions of wavelength $\lambda$ of the square lattice array of Ag nanocubes on top of 5 nm of hBN on graphene heterostructure shown in Fig.~\ref{fig:AgNanocube-graphene}. Graphene on the 1400 nm thick FeCl$_3$ spacer has a mobility of $\mu=1000$ V/cm$^2$s. A gate voltage is applied to graphene that varies the Fermi energy $E_F=-1.0,-0.9,-0.8,-0.7,-0.6,-0.5,-0.4$ eV. The Ag nanocube side length is $a=b=c=70$ nm. The period of the square lattice is $\calP=90$ nm.  The results for $I(\lambda)$ are obtained by FDTD calculations for Fermi energies $E_F=-1.0,-0.8,-0.6,-0.4$ eV at $T=300,310,320$ K. The resonance peak of the AGP can be tuned by means of the gate voltage in the wavelength regime between $\lambda=5$ $\mu$m and $\lambda=8$ $\mu$m for the radiative heat trapping mode, and in the wavelength regime between $\lambda=8$ $\mu$m and $\lambda=12$ $\mu$m for the radiative cooling mode.
\label{fig:AgNanocube-graphene_emittance} }
\end{figure*}

Considering the boundary conditions of the graphene sheet and its image [see Eqs.~(\ref{eq:graphene_boundary}) and (\ref{eq:graphene_image_boundary}), respectively] we obtain the following equations:
\begin{align}
    E_{x<}(z=-h) &=i\frac{\zeta_{z<} k_x}{\omega\mu\epsilon}A_{<}\cos\left(k_x x\right)\sin\left(k_y y\right)e^{-\zeta_{z<} h}, \nn\\
    || \hspace{1cm} & \nn\\
    E_{xmn}(z=-h) &=i\frac{\zeta_{zmn} k_x}{\omega\mu\epsilon}A_{mn}\cos\left(k_x x\right)\sin\left(k_y y\right)\sinh\left(-\zeta_{zmn} h\right), \nn\\
    H_{y<}(z=-h) &= -\frac{k_y}{\mu}A_{<}\sin\left(k_x x\right)\cos\left(k_y y\right)e^{-\zeta_{z<} h}, \nn\\
    || \hspace{1cm} & \nn\\
    H_{ymn}(z=-h) &+ \mu_0\sigma E_{x<}(z=-h) = -\frac{k_y}{\mu}A_{mn}\sin\left(k_x x\right)\cos\left(k_y y\right)\cosh\left(-\zeta_{zmn} h\right) \nn\\
    &+ \mu_0\sigma i\frac{\zeta_{zmn} k_x}{\omega\mu\epsilon}A_{<}\cos\left(k_x x\right)\sin\left(k_y y\right)e^{-\zeta_{z<} h}
\end{align}
where $<$ is an abbreviation for $z\le -h$. Similar equations hold for $z=h$, where $<$ needs to be replaced by $>$, an abbreviation for $z\ge h$.
The short form can be written as
\begin{align}
    E_{x<0}e^{-\zeta_{z<} h} &= E_{xmn0}\sinh\left(-\zeta_{zmn} h\right), \nn\\
    H_{y<0}e^{-\zeta_{z<} h} &= H_{ymn0}\cosh\left(-\zeta_{zmn} h\right)+\mu_0\sigma E_{x<0}e^{-\zeta_{z<} h}  \nn\\
    E_{x>0}e^{-\zeta_{z>} h} &= E_{xmn0}\sinh\left(\zeta_{zmn} h\right), \nn\\
    H_{y>0}e^{-\zeta_{z>} h} &= H_{ymn0}\cosh\left(\zeta_{zmn} h\right)+\mu_0\sigma E_{x>0}e^{-\zeta_{z>} h}
\end{align}
Using Maxwell's equations in dielectric media, from which one gets ${\rm sgn}(z)\zeta_z H_y = -i(\omega\epsilon_r/c^2)E_x$, we obtain\cite{Goncalves2012}

\begin{align}
    E_{x<0}e^{-\zeta_{z<} h} &= E_{xmn0}\sinh\left(-\zeta_{zmn} h\right), \nn\\
    i\frac{\omega\epsilon_r}{c^2\zeta_{z<}}E_{x<0}e^{-\zeta_{z<} h}
    &= i\frac{\omega\epsilon_r}{c^2\zeta_{zmn}}E_{xmn0}\cosh\left(-\zeta_{zmn} h\right)+\mu_0\sigma E_{x<0}e^{-\zeta_{z<} h}  \nn\\
    E_{x>0}e^{-\zeta_{z>} h} &= E_{xmn0}\sinh\left(\zeta_{zmn} h\right), \nn\\
    -i\frac{\omega\epsilon_r}{c^2\zeta_{z>}}E_{x>0}e^{-\zeta_{z>} h} 
    &= -i\frac{\omega\epsilon_r}{c^2\zeta_{zmn}}E_{xmn0}\cosh\left(\zeta_{zmn} h\right)+\mu_0\sigma E_{x>0}e^{-\zeta_{z>} h}
\end{align}
Because of the mirror symmetry there are only two independent variables. The equations above can be written in terms of a matrix equation as
\be
    \bmaa 
        -e^{-\zeta_{z>} h} & \sinh\left(\zeta_{zmn} h\right) \\
        \left[\mu_0\sigma+i\frac{\omega\epsilon_r}{c^2\zeta_{z>}}\right]e^{-\zeta_{z>} h} & -i\frac{\omega\epsilon_r}{c^2\zeta_{zmn}}\cosh\left(\zeta_{zmn} h\right)
    \ema 
    \bma
        E_{x>0} \\
        E_{xmn0}
    \ema 
    =0
\ee
This matrix equation has nontrivial solutions only if ${\rm det}M=0$, where $M$ is the matrix in the above equation. One obtains\cite{Goncalves2012}
\begin{align}
    i\frac{\omega\epsilon_r}{c^2\zeta_{zmn}}\cosh\left(\zeta_{zmn} h\right)e^{-\zeta_{z>} h}
    &= \left[\mu_0\sigma+i\frac{\omega\epsilon_r}{c^2\zeta_{z>}}\right]\sinh\left(\zeta_{zmn} h\right)e^{-\zeta_{z>} h} \nn\\
    \Leftrightarrow
    i\frac{\omega\epsilon_r}{c^2\zeta_{zmn}}
    &= \left[\mu_0\sigma+i\frac{\omega\epsilon_r}{c^2\zeta_{z>}}\right]\tanh\left(\zeta_{zmn} h\right) \nn\\
    \Leftrightarrow
    1 &= \left[-i\frac{\mu_0\sigma c^2\zeta_{zmn}}{\omega\epsilon_r}
    +\frac{\zeta_{zmn}}{\zeta_{z>}}\right]\tanh\left(\zeta_{zmn} h\right) \nn\\
    \Leftrightarrow
    \coth\left(\zeta_{zmn} h\right) &= -i\frac{\mu_0\sigma c^2\zeta_{zmn}}{\omega\epsilon_r}+
    \frac{\zeta_{zmn}}{\zeta_{z>}} \nn\\
    \Leftrightarrow
    \frac{\epsilon_r}{\zeta_{zmn}}\coth\left(\zeta_{zmn} h\right) &= -i\frac{\sigma}{\epsilon_0\omega}
    +\frac{\epsilon_r}{\zeta_{z>}}.
    \label{eq:acoustic_mode_equation}
\end{align}
This is the acoustic graphene plasmon (AGP) polariton mode, which is allowed by the mirror symmetry.
As a side note, one would have to choose the ansatz $A_z\propto \sinh\left(\zeta_z z\right)$ in Eq.~(\ref{eq:A_ansatz_graphene}) to obtain the optical plasmon polariton mode. However, it is not allowed by the mirror symmetry.

\begin{figure}[h]
\begin{centering}
\includegraphics[width=16cm]{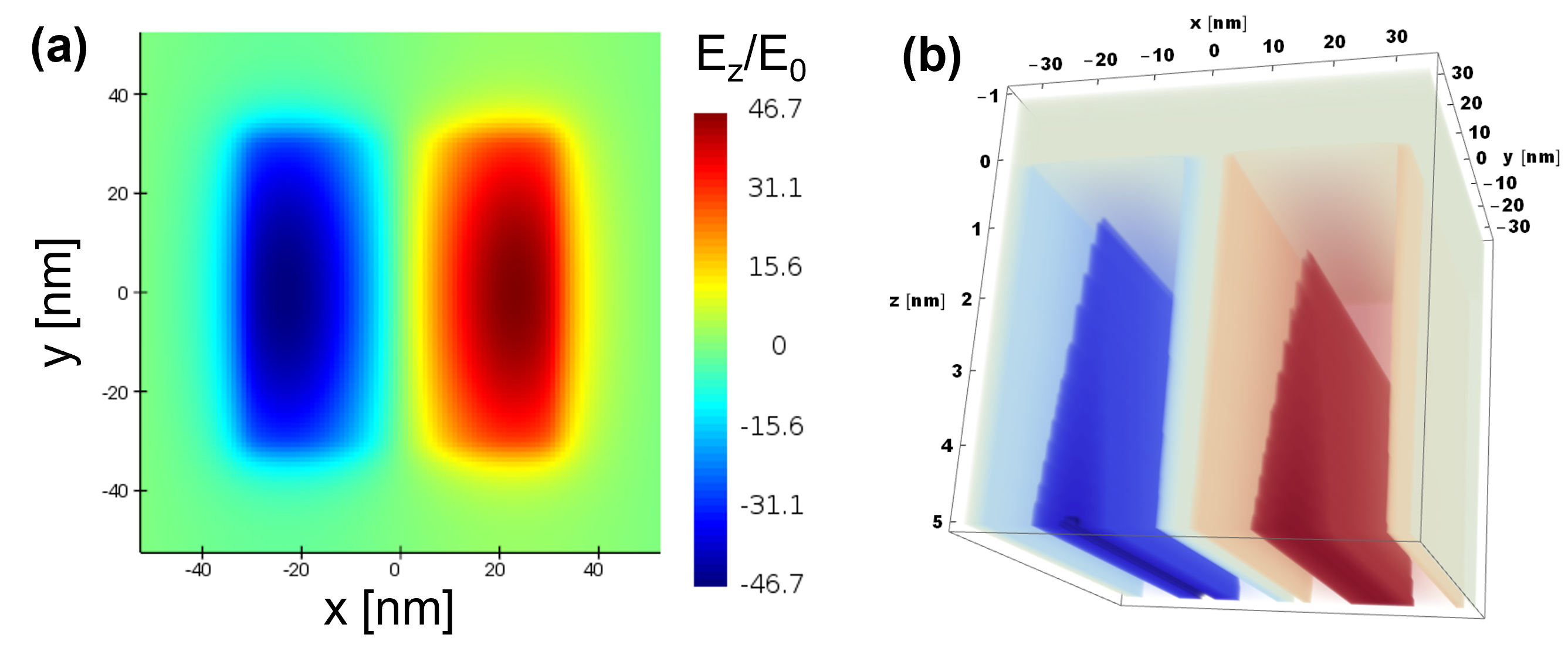}
\end{centering}
\caption{$E_z$ field of an acoustic graphene plasmon (AGP) in a metal-dielectric-graphene square cavity. The graphene sheet is at the bottom, the metal nanocube at the top. The volume $h\times L\times W=5$ nm $\times 60$ nm $\times 60$ nm is occupied by the dielectric. (a) FDTD result and (b) analytical result for a graphene sheet at $E_F=1.0$ eV, corresponding to the AGP resonance peak at $\lambda=5.8$ $\mu$m in Fig.~\ref{fig:AgNanocube-graphene_emittance}. The analytical result is in good agreement with the FDTD result. The $E_z$ field component of the AGP solution is clearly localized beneath the Ag nanocube
\label{fig:Ez_nanocube-dielectric-graphene} }
\end{figure}

In the case of ${\varepsilon _F} \gg {k_B}T$ the intraband optical conductivity in graphene is
 \be
\sigma_{\rm intra}(\omega) = \frac{e^2}{\pi\hbar^2}\frac{E_F}{\tau^{-1} - i\omega }=\frac{2\varepsilon_m\omega_p^2}{\pi\hbar^2(\tau^{-1}-i\omega)},
\label{eq:sigma_intra}
 \ee
 where $\tau$ is determined by impurity scattering and electron-phonon interaction ${\tau ^{ - 1}} = \tau _{imp}^{ - 1} + \tau _{e - ph}^{ - 1}$ .
 Using the mobility $\mu$ of the graphene sheet, it can be presented in the form
 $\tau^{-1}=ev_F^2/(\mu E_F)$, where $v_F=10^6$ m/s is the Fermi velocity in graphene.
 $\omega_p=\sqrt{e^2E_F/2\varepsilon_m}$ is the bulk graphene plasma frequency.
 
\begin{figure}[h]
\begin{centering}
\includegraphics[width=16cm]{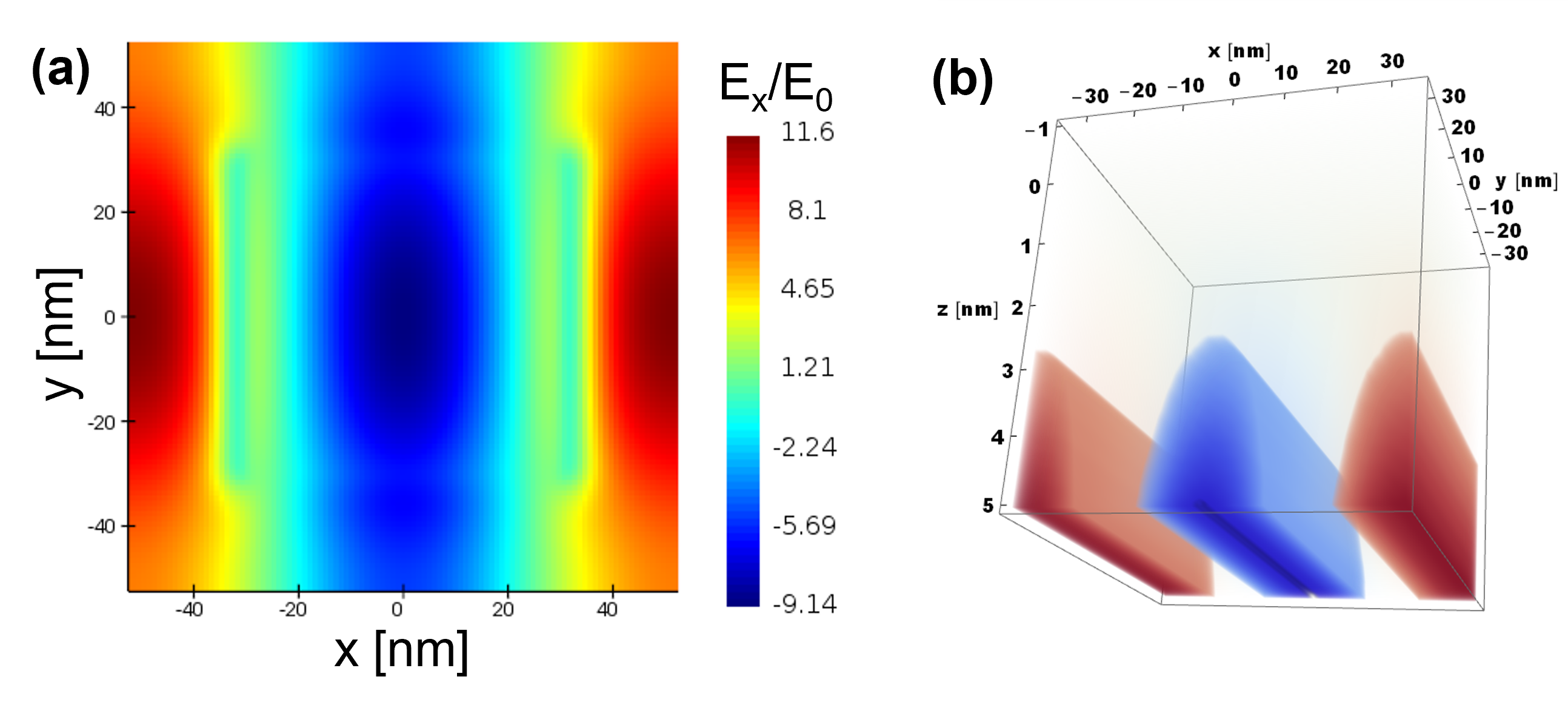}
\end{centering}
\caption{$E_x$ field of an acoustic graphene plasmon (AGP) in a metal-dielectric-graphene square cavity. The graphene sheet is at the bottom, the metal nanocube at the top. The volume $h\times L\times W=5$ nm $\times 60$ nm $\times 60$ nm is occupied by the dielectric. (a) FDTD result and (b) analytical result for a graphene sheet at $E_F=1.0$ eV, corresponding to the AGP resonance peak at $\lambda=5.8$ $\mu$m in Fig.~\ref{fig:AgNanocube-graphene_emittance}. This result is in stark contrast to the $E_x$ field in the case of the metal patch-dielectric-metal nanostructure, for which $E_x=0$.
The analytical result is in good qualitative agreement with the FDTD result. The main difference is that the analytical result is localized beneath the Ag nanocube whereas the FDTD result shows coupling between the Ag nanocubes, i.e. the $E_x$ field is located also in the space between the Ag nanocubes. This means that the $E_x$ field component of the AGP solution is delocalized.
\label{fig:Ex_nanocube-dielectric-graphene} }
\end{figure}

Inserting this intraband optical condutivity into Eq.~(\ref{eq:acoustic_mode_equation}), we obtain
\be
    \frac{\epsilon_r}{\zeta_{zmn}}\coth\left(\zeta_{zmn} h\right) 
    = \frac{e^2}{\pi\hbar^2}\frac{E_F}{\epsilon_0\omega\left(\omega+i\tau^{-1}\right)}
    +\frac{\epsilon_r}{\zeta_{z>}}.
\ee
Assuming $\zeta_{zmn} h\ll 1$, we can approximate $\coth\left(\zeta_{zmn} h\right)\approx 1/(\zeta_{zmn} h)$, which simplifies the above equation to
\be
    \left(\frac{1}{\zeta_{zmn}^2 h}- \frac{1}{\zeta_{z>}} \right)
    = \frac{e^2}{\pi\hbar^2}\frac{E_F}{\epsilon_0\epsilon_r\omega\left(\omega+i\tau^{-1}\right)}.
\ee
Assuming $\zeta_{zmn} \ll \zeta_{z>}$, we obtain
\be
    \frac{1}{\zeta_{zmn}^2 h}
    = \frac{e^2}{\pi\hbar^2}\frac{E_F}{\epsilon_0\epsilon_r\omega\left(\omega+i\tau^{-1}\right)}.
\ee
Next we use $\zeta_{zmn}^2=q_{zmn}^2-\epsilon_r\omega^2/c^2$ with $q_{zmn}^2=k_{xm}^2+k_{yn}^2$ to get
\begin{align}
    \omega\left(\omega+i\tau^{-1}\right) &= \frac{e^2}{\pi\hbar^2}
    \frac{h E_F\left(q_{zmn}^2-\epsilon_r\omega^2/c^2\right)}{\epsilon_0\epsilon_r} \nn\\
    \Leftrightarrow
    \left(1+\frac{e^2 h E_F}{\pi\hbar^2\epsilon_0 c^2}\right)\omega^2 
    +i\tau^{-1}\omega 
    &= \frac{e^2}{\pi\hbar^2}
    \frac{h E_F\left(q_{zmn}^2\right)}{\epsilon_0\epsilon_r} .
\end{align}
The graphene SPP solutions are
\begin{align}
    \omega & = \frac{-i\tau^{-1}\pm \sqrt{-\tau^{-2}
    +4\left(1+\frac{e^2 h E_F}{\pi\hbar^2\epsilon_0 c^2}\right)\frac{e^2}{\pi\hbar^2}
    \frac{h E_F\left(q_{zmn}^2\right)}{\epsilon_0\epsilon_r}}}{2\left(1+\frac{e^2 h E_F}{\pi\hbar^2\epsilon_0 c^2}\right)} \nn\\
    &\approx \frac{-i\tau^{-1}\pm 
    \sqrt{4\left(1+\frac{e^2 h E_F}{\pi\hbar^2\epsilon_0 c^2}\right)\frac{e^2}{\pi\hbar^2}
    \frac{h E_F\left(q_{zmn}^2\right)}{\epsilon_0\epsilon_r}}}{2\left(1+\frac{e^2 h E_F}{\pi\hbar^2\epsilon_0 c^2}\right)} \nn\\
    &= \frac{-i\tau^{-1}}{2\left(1+\frac{e^2 h E_F}{\pi\hbar^2\epsilon_0 c^2}\right)}
    \pm 
    q_{zmn}\sqrt{\frac{\frac{e^2}{\pi\hbar^2}\frac{h E_F}{\epsilon_0\epsilon_r}}{\left(1+\frac{e^2 h E_F}{\pi\hbar^2\epsilon_0 c^2}\right)}} \nn\\
    &\approx -i\tau^{-1}/2 \pm q_{zmn}\sqrt{\frac{e^2}{\pi\hbar^2}\frac{h E_F}{\epsilon_0\epsilon_r}} \nn\\
    &= \pm 2q_{zmn}\sqrt{\alpha\frac{c h E_F}{\hbar\epsilon_r}} -i\tau^{-1}/2,
\end{align}
where in the last approximation we used $\frac{e^2 h E_F}{\pi\hbar^2\epsilon_0 c^2}\ll 1$.
The fine structure constant is defined as $\alpha=e^2/(4\pi\epsilon_0\hbar c)$.
Note the linear dispersion for this AGP polariton mode!

\begin{figure*}[htb]
\begin{centering}
\includegraphics[width=16.0cm]{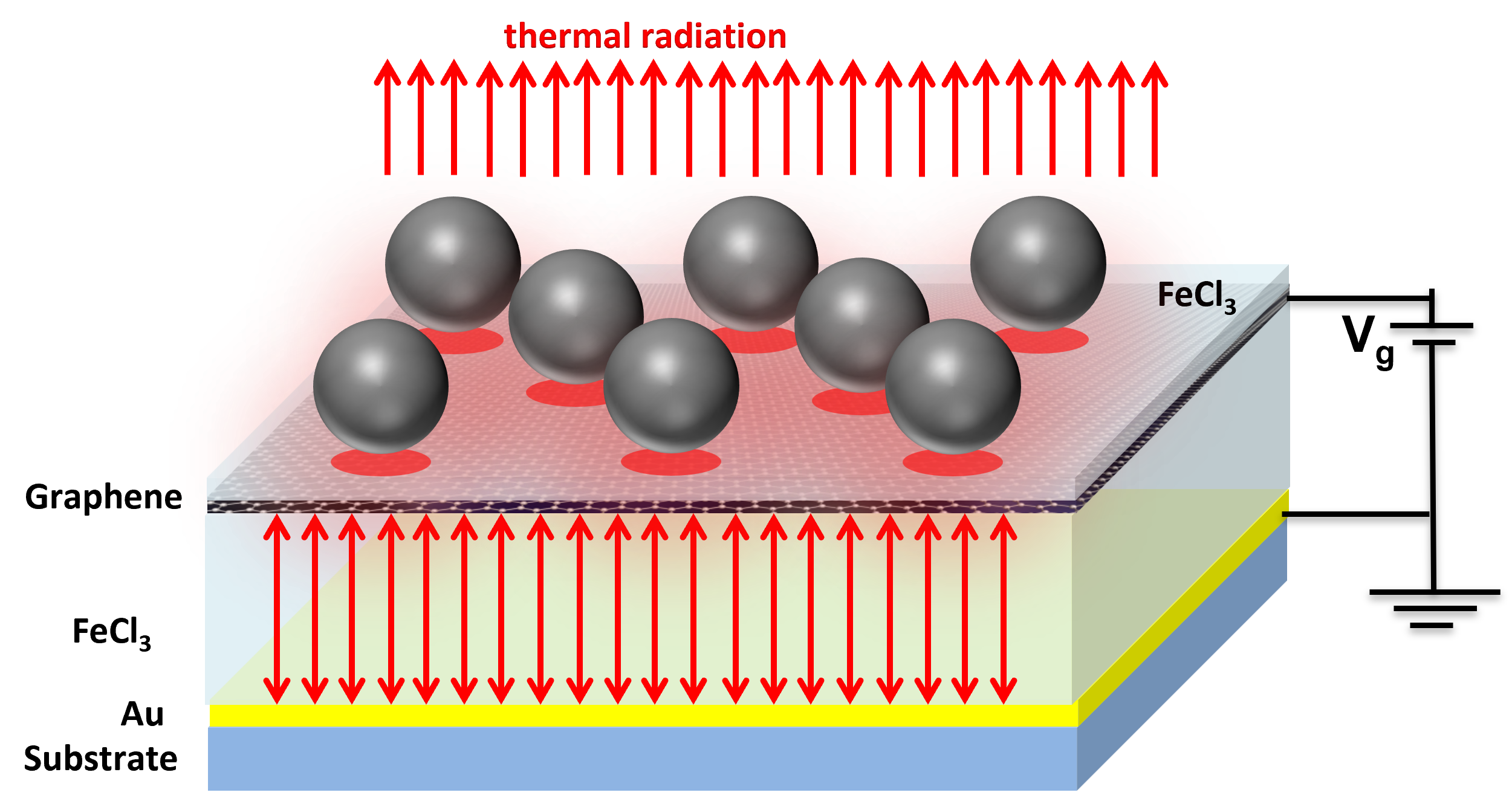}
\end{centering}
\caption{Schematic showing our proposed Ag nanosphere/hBN/graphene heterostructure placed on top of a cavity, which can be tuned by means of a gate voltage applied to the Au back mirror. The spacer of the cavity consists of FeCl$_3$.  
\label{fig:AgNanosphere-graphene} }
\end{figure*}

Thus, the resonant frequencies of the metal-dielectric-graphene cavity are
\begin{align}
    f_{\rm AGP} &= \sqrt{\alpha\frac{c h E_F}{\pi^2\hbar\epsilon_r}}
    \sqrt{\left(\frac{m\pi}{L}\right)^2+\left(\frac{n\pi}{W}\right)^2} \nn\\
    &= \frac{v_{\rm AGP}}{2\pi}\sqrt{\left(\frac{m\pi}{L}\right)^2+\left(\frac{n\pi}{W}\right)^2},
\end{align}
where we defined 
\be
v_{\rm AGP}=2\sqrt{\alpha\frac{c h E_F}{\hbar\epsilon_r}}
\ee 
as the effective phase velocity of the AGP mode inside the cavity.
Remarkably, $v_{\rm AGP}$ can be tuned by means of the Fermi energy $E_F$ in graphene and the thickness $h$ of the dielectric between the metal and the graphene sheet.

 \begin{figure*}[htb]
\begin{centering}
\includegraphics[width=16cm]{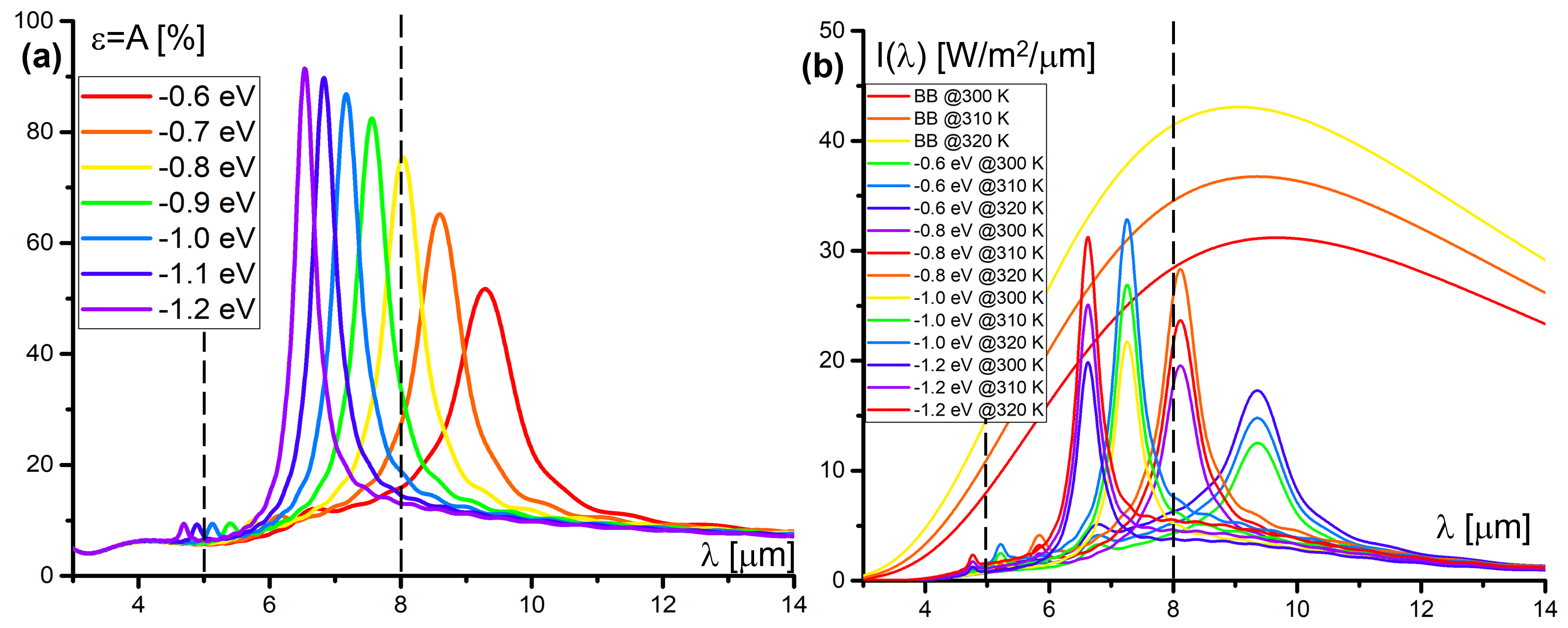}
\end{centering}
\caption{(a) Emittance $\epsilon(\lambda)$ (=absorbance $A(\lambda)$) and (b) spectral radiance $I(\lambda)$ as functions of wavelength $\lambda$ of the hexagonal lattice array of Ag nanospheres on top of 5 nm of hBN on graphene heterostructure shown in Fig.~\ref{fig:AgNanosphere-graphene}. Graphene on the 900 nm thick FeCl$_3$ spacer has a mobility of $\mu=1000$ V/cm$^2$s. A gate voltage is applied to graphene that varies the Fermi energy $E_F=-1.0,-0.9,-0.8,-0.7,-0.6$ eV. The Ag nanosphere radius is $R=130$ nm. The period of the hexagonal lattice is $\calP=176$ nm.  The results for $I(\lambda)$ are obtained by FDTD calculations for Fermi energies $E_F=-1.2,-1.0,-0.8,-0.6$ eV at $T=300,310,320$ K. The resonance peak of the AGP can be tuned by means of the gate voltage in the wavelength regime between $\lambda=5$ $\mu$m and $\lambda=8$ $\mu$m for radiative heat trapping mode, and in the wavelength regime between $\lambda=8$ $\mu$m and $\lambda=12$ $\mu$m for radiative cooling mode.
\label{fig:AgNanosphere-graphene_emittance} }
\end{figure*}

\subsection{Results and Discussion}
The finite-difference time domain (FDTD) results for the emittance $\epsilon(\lambda)$ and the spectral radiance $I(\lambda)$ shown in Fig.~\ref{fig:AgNanocube-graphene_emittance} at $T=300,310,320$ K demonstrate spectrally selective thermal emission from Ag nanocubes on top of hBN/graphene at a wavelength that corresponds to the AGP resonance. The Ag nanocube side length is $a=b=c=70$ nm. The vertices of the nanocubes are slightly rounded in the FDTD simulations, in agreement with commercially available nanocubes.\cite{SigmaAldrich_nanocubes} The period of the square lattice is $\calP=105$ nm. The hBN layer between the Ag nanocube and graphene is $h=5$ nm thick. The FeCl$_3$ spacer is $1400$ nm thick. The AGP resonance peak can be shifted between $\lambda=5$ $\mu$m and $\lambda=12$ $\mu$m by means of a gate voltage $V_g$ that tunes continuously the Fermi energy between $E_F=-0.4$ eV and $E_F=-1.0$ eV. Remarkably, the emittance is nearly 100\% for larger values of the Fermi energies.
The spectral radiance is calculated according to
\be
I(\omega)=\epsilon(\omega)\frac{\omega^2}{4\pi^3c^2}\Theta(\omega,T),
\ee
where $\omega=2\pi c/\lambda$ is the angular frequency of the emitted photon, $c$ is the vacuum speed of light, and $\Theta(\omega,T)=\hbar\omega/[\exp(\hbar\omega/k_BT)-1]$ is the thermal energy of a photon mode. $I(\omega)$ is calculated for Fermi energies of $E_F=-1.0,-0.8,-0.6, -0.4$ eV at $T=300,310,320$ K.
According to the FDTD result of the electric field profiles $E_z(x,y)$ and $E_x(x,y)$ shown in Figs.~\ref{fig:Ez_nanocube-dielectric-graphene} and \ref{fig:Ex_nanocube-dielectric-graphene}, respectively, we can identify by means of the analytical result the AGP mode $m=2$, $n=0$. The resonance wavelength of the analytical result is $\lambda=6.8$ $\mu$m, which is in good agreement with the resonance wavelength from the FDTD result $\lambda=6.6$ $\mu$m for $E_F=-1.0$ eV.

\section{AGP for the nanosphere-dielectric-graphene system}
\begin{figure}[h]
\begin{centering}
\includegraphics[width=16cm]{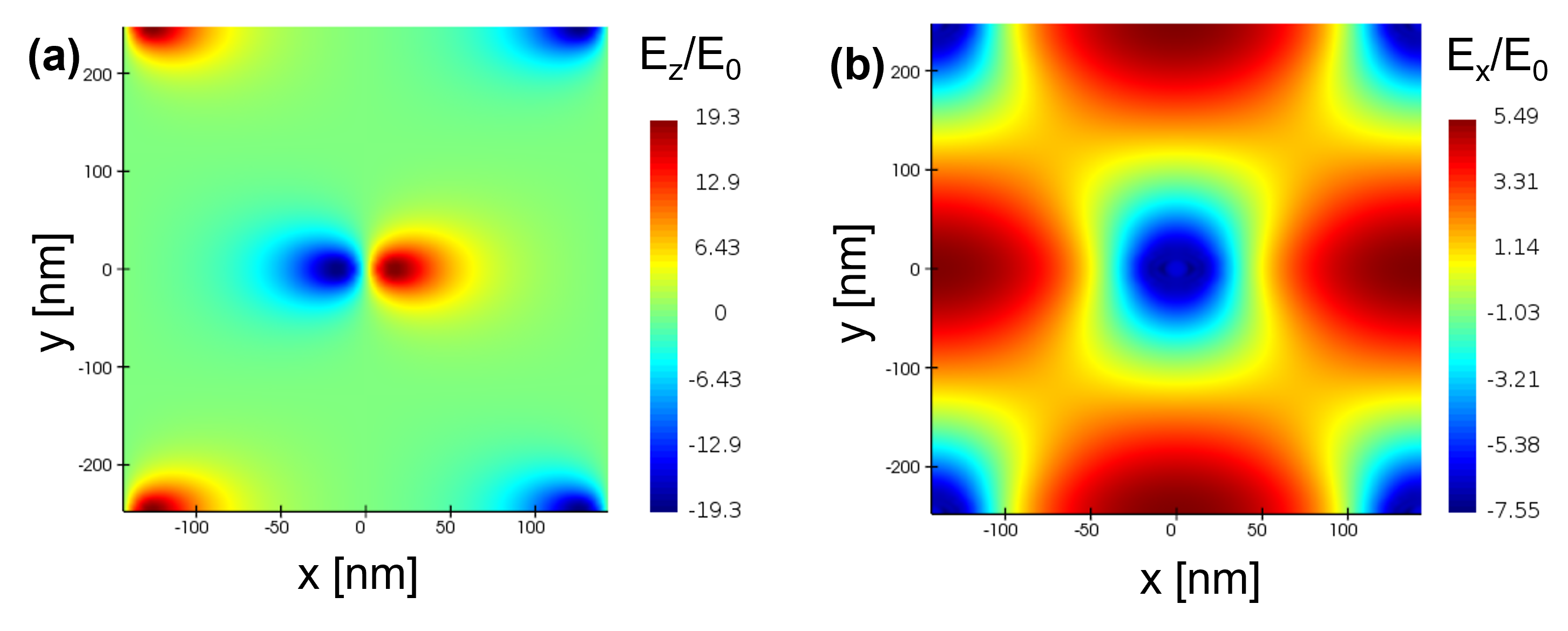}
\end{centering}
\caption{(a) $E_z$ field and (b) $E_x$ field components of an acoustic graphene plasmon (AGP) in a metal-dielectric-graphene sphere cavity. The graphene sheet is at the bottom, the metal nanosphere at the top. The $E_z$ field component of the AGP solution is dipolar in nature and clearly localized beneath the Ag nanosphere.
\label{fig:Ez&Ex_nanosphere-dielectric-graphene} }
\end{figure}

Next we consider nanospheres on top of hBN/graphene, as shown in Fig.~\ref{fig:AgNanosphere-graphene}. In contrast to the case of the Ag nanocubes, for which the square lattice results in maximum emittance, we find that for the Ag nanospheres the hexagonal lattice gives maximum emittance. We perform FDTD calculations to obtain the AGP resonance peaks and the AGP electric field profiles.
The FDTD results for the emittance $\epsilon(\lambda)$ and the spectral radiance $I(\lambda)$ shown in Fig.~\ref{fig:AgNanosphere-graphene_emittance} at $T=300,310,320$ K demonstrate spectrally selective thermal emission from Ag nanospheres on top of hBN/graphene at a wavelength that corresponds to the AGP resonance. The Ag nanosphere radius is $R=130$ nm. The period of the hexagonal lattice is $\calP=286$ nm. The hBN layer between the Ag nanosphere and graphene is $h=5$ nm thick. The FeCl$_3$ spacer is $900$ nm thick. The AGP resonance peak can be shifted between $\lambda=5$ $\mu$m and $\lambda=12$ $\mu$m by means of a gate voltage $V_g$ that tunes continuously the Fermi energy between $E_F=-0.6$ eV and $E_F=-1.2$ eV. $I(\omega)$ is calculated for Fermi energies of $E_F=-1.2, -1.0,-0.8,-0.6$ eV at $T=300,310,320$ K.
The FDTD result of the electric field profiles $E_z(x,y)$ and $E_x(x,y)$ are shown in Fig.~\ref{fig:Ez&Ex_nanosphere-dielectric-graphene}. 
The $E_z$ field component of the AGP is clearly dipolar in nature. The $E_x$ field component of the AGP is delocalized, similarly to the case of the AGP for Ag nanocubes.

\section{Conclusion}
In conclusion, we have demonstrated in our theoretical study that metallic nanoparticles on graphene can be used to develop a surface for spectrally selective thermal emission in the IR regime between $\lambda=5$ $\mu$m and $\lambda=12$ $\mu$m by means of the AGP modes trapped inside cavities between the metallic nanoparticles and graphene.
Most importantly, the AGPs along with an optical cavity increase substantially the emittance (and therefore also the absorbance) of graphene from about 2\% for pristine graphene to nearly 100\% for metallic nanoparticles on graphene, thereby outperforming state-of-the-art graphene IR light sources working in the visible and NIR by at least a factor of 100.

The spectrally selective thermal emission from Ag nanoparticles on top of graphene will pave the way to develop nanostructured graphene-based fabrics for winter and summer clothes that improve the thermal management of the human body, electronic devices, and batteries in cold and warm temperature environments. Being able to tune the spectrally selective thermal emission in the atmospherically opaque IR bandwidth between $\lambda=5$ $\mu$m and $\lambda=8$ $\mu$m and also in the atmospherically transparent IR bandwidth between $\lambda=8$ $\mu$m and $\lambda=12$ $\mu$m allows for electric switching between radiative heat trapping mode and radiative cooling mode.

\section{Methods}
The optical FDTD simulations have been performed using Ansys Lumerical FDTD software package. We use the evaporated gold optical model described in Palik et. al.\cite{palik1998handbook}. We model graphene as a thin conductive layer using the optical conductivity of graphene. The mobility and Fermi energy dependent scattering rate ($\tau^{-1}=ev_F^2/(\mu E_F)$) for graphene is used to calculate absorption spectra for different Fermi levels. The mesh size between the Ag nanoparticles and graphene is set to 1 nm in the xy plane, and to 0.05 nm in the z direction, with an auto minimum shutoff of $10^{-6}$ and simulation time of 5000 fs.

 

\begin{acknowledgement}

M.N.L. and D.R.E. acknowledge support by DARPA/DSO under grant no. HR00112220011.
M.N.L. achknowledges support by the ORISE fellowship 2022.

\end{acknowledgement}

\bibliography{Graphene_IR_detector}

\end{document}